\documentclass[11pt,a4paper,useAMS,usenatbib]{emulateapj}
\usepackage{epsfig}
\usepackage{amsmath}
\usepackage{natbib}
\usepackage{rotating}

\begin{document}

\title{Probing the Hot X-ray Corona around the Massive Spiral Galaxy, NGC~6753, \\ Using Deep XMM-Newton observations}

\author{\'Akos Bogd\'an\altaffilmark{1}, Herv\'e Bourdin\altaffilmark{1}, William R. Forman\altaffilmark{1}, Ralph P. Kraft\altaffilmark{1}, \\ Mark Vogelsberger\altaffilmark{2}, Lars Hernquist\altaffilmark{1}, and Volker Springel\altaffilmark{3,4}}
\affil{\altaffilmark{1}Harvard-Smithsonian Center for Astrophysics, 60 Garden Street, Cambridge, MA 02138, USA}
\affil{\altaffilmark{2}Department of Physics, Kavli Institute for Astrophysics and Space Research, Massachusetts Institute of Technology, Cambridge, MA 02139, USA}
\affil{\altaffilmark{3}Heidelberg Institute for Theoretical Studies, Schloss-Wolfsbrunnenweg 35, D-69118 Heidelberg, Germany}
\affil{\altaffilmark{4}Zentrum f\"ur Astronomie der Universit\"at Heidelberg, ARI, M\"onchhofstr. 12-14, D-69120 Heidelberg, Germany}

\email{E-mail: abogdan@cfa.harvard.edu}

\shorttitle{HOT X-RAY CORONA AROUND NGC~6753}
\shortauthors{BOGD\'AN ET AL.}

\begin{abstract}
X-ray emitting gaseous coronae around massive galaxies are a basic prediction of galaxy formation models. Although the coronae around spiral galaxies offer a fundamental test of these models, observational constraints on their characteristics are still scarce. While the presence of extended hot coronae has been established around a handful of massive spiral galaxies, the short X-ray observations only allowed for measurements of the basic characteristics of the coronae. In this work, we utilize deep \textit{XMM-Newton} observations of NGC~6753 to explore its extended X-ray corona in unprecedented detail. Specifically, we establish the isotropic morphology of the hot gas, suggesting that it resides in hydrostatic equilibrium. The temperature profile of the gas shows a decrease with increasing radius: it drops from $kT\approx0.7$ keV in the innermost parts to $kT\approx0.4$ keV at 50 kpc radius. The temperature map reveals the complex temperature structure of the gas. We study the metallicity distribution of the gas, which is uniform at $Z\approx0.1$ Solar. This value is about an order of magnitude lower than that obtained for elliptical galaxies with similar dark matter halo mass, hinting that the hot gas in spiral galaxies predominantly originates from external gas inflows rather than from internal sources. By extrapolating the density profile of the hot gas out to the virial radius, we estimate the total gas mass and derive the total baryon mass of NGC 6753. We conclude that the baryon mass fraction is $f_{\rm b} \approx 0.06$, implying that about half of the baryons are missing.
\end{abstract}

\keywords{galaxies: individual (NGC~6753)  --- galaxies: spiral --- galaxies: ISM  --- X-rays: galaxies --- X-rays: general --- X-rays: ISM}

\section{Introduction}
\label{sec:introduction}
According to every modern galaxy formation model, massive galaxies in the present-day universe are expected to host quasi-static luminous X-ray coronae \citep[e.g.][]{white78,white91,benson00,toft02,crain10,vogelsberger12}. The gaseous coronae -- which extend out to the virial radius of galaxies -- are believed to have emerged in the earliest epochs of galaxy formation. In the spherical infall model, the gas falling onto the dark matter halos is heated by accretion shocks to the virial temperature of the dark matter halo \citep{larson74}. For massive halos ($M_{\rm vir} \gtrsim 10^{12} \ \rm{M_{\rm \odot}}$), the virial temperature exceeds $\sim100$ eV, hence the shock heated gas is expected to emit X-rays. In the innermost regions the density of the gas is high and it can radiatively cool and supply star formation in the disk of the galaxy. However, predominant fraction of the gas residing at large radii has low density, implying that the cooling time of the gas is comparable or longer than the Hubble-time, resulting in the formation of quasi-static X-ray coronae.  Given that these coronae were accreted and heated in the early universe, studying them provides a direct insight into the most fundamental processes of galaxy formation. 

Although galaxy formation models predict extended X-ray coronae around galaxies with all morphological types, the cleanest tests are provided by relatively isolated disk-dominated (spiral) galaxies with quiescent star formation histories. Unlike their elliptical counterparts, massive spiral galaxies are often found in isolation and not in the center of rich galaxy groups and clusters that have their own gaseous coronae. In addition, spiral galaxies did not undergo major galaxy mergers, which may trigger vigorous star formation activity and the quasar phase of the central supermassive black hole, thereby altering the characteristics of the primordial coronae. Although massive spiral galaxies are ideal to probe galaxy formation models, due to the scarcity of such galaxies and the faint X-ray luminosities, it is difficult to explore their coronae in detail.  

\begin{figure*}[t]
  \begin{center}
    \leavevmode
      \epsfxsize=0.98\textwidth\epsfbox{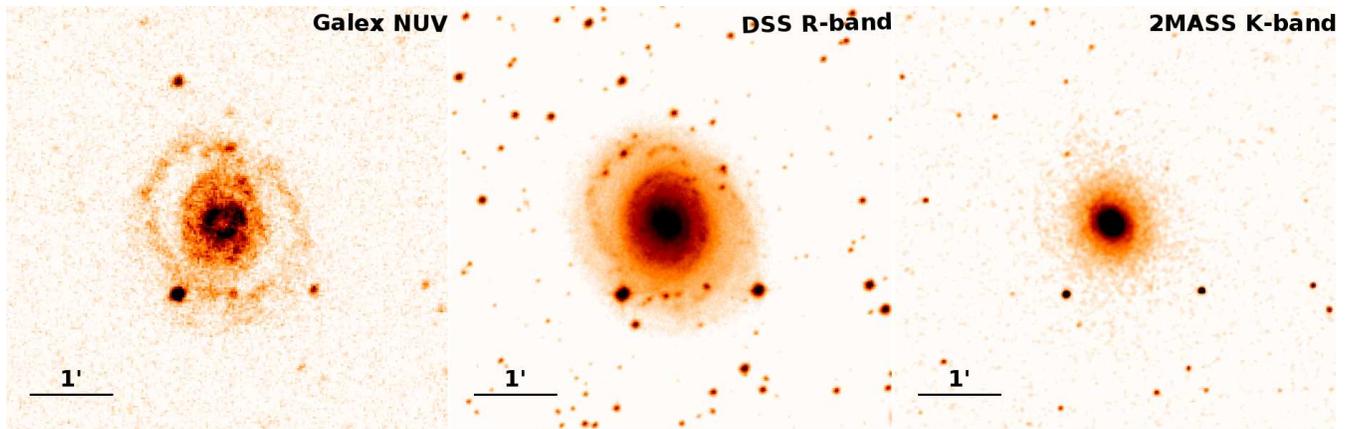}
\vspace{0.2cm}
      \caption{Multi-wavelength images of NGC~6753. The star-formation activity is traced by the near-ultraviolet image taken by \textit{Galex} (left panel). The image reveals the presence of two star-forming rings: the inner one at about $0.2\arcmin$ ($\approx2.5$ kpc) and the outer one at $1\arcmin$ ($\approx12.5$ kpc) from the center. The \textit{DSS} R-band image shows the stellar light distribution (middle panel). The star-formation activity seen on the \textit{Galex} near-ultraviolet image is associated with the spiral arms of the galaxy. The \textit{2MASS} K-band image traces the old stellar population, hence the K-band image is more compact than that of the R-band image and the bulge of NGC~6753 is the most pronounced. All three images reveal that the stellar light and the associated star-formation does not extend beyond $1.2\arcmin$ ($\approx15$ kpc).}
     \label{fig:multi}
  \end{center}
\end{figure*}

Given the fundamental nature of these coronae, all major X-ray observatories of the past decades attempted to explore the X-ray coronae around massive spiral galaxies. These attempts, mostly focusing on nearby edge-on disk galaxies, remained unsuccessful for decades \citep[e.g.][]{wang05,li08,rasmussen09,li11,bogdan15}. The breakthrough has been achieved recently, when \textit{XMM-Newton}, \textit{Chandra}, and \textit{ROSAT} observations revealed the presence of extended coronae around four massive spiral galaxies  \citep{anderson11,dai12,bogdan13a,bogdan13b}. The  importance of these detections are immense; they demonstrate that relaxed spiral galaxies have gaseous X-ray coronae that extend well beyond the stellar light and -- in a more general context -- note that our basic picture of galaxy formation is correct. However, it was possible to characterize the hot gas in only two galaxies: NGC~6753 and NGC~1961. Due to the relatively short initial observations, only the average properties, namely the gas temperature and the metallicity, of the gas could be established. Therefore, the detailed temperature and metallicity structure remained unexplored. For the other two detected galaxies (NGC~266 and UGC~12591) only the most basic properties of the gas, such as its extent and approximate luminosity, could be established. 

\begin{figure*}[t]
  \begin{center}
    \leavevmode
      \epsfxsize=0.8\textwidth\epsfbox{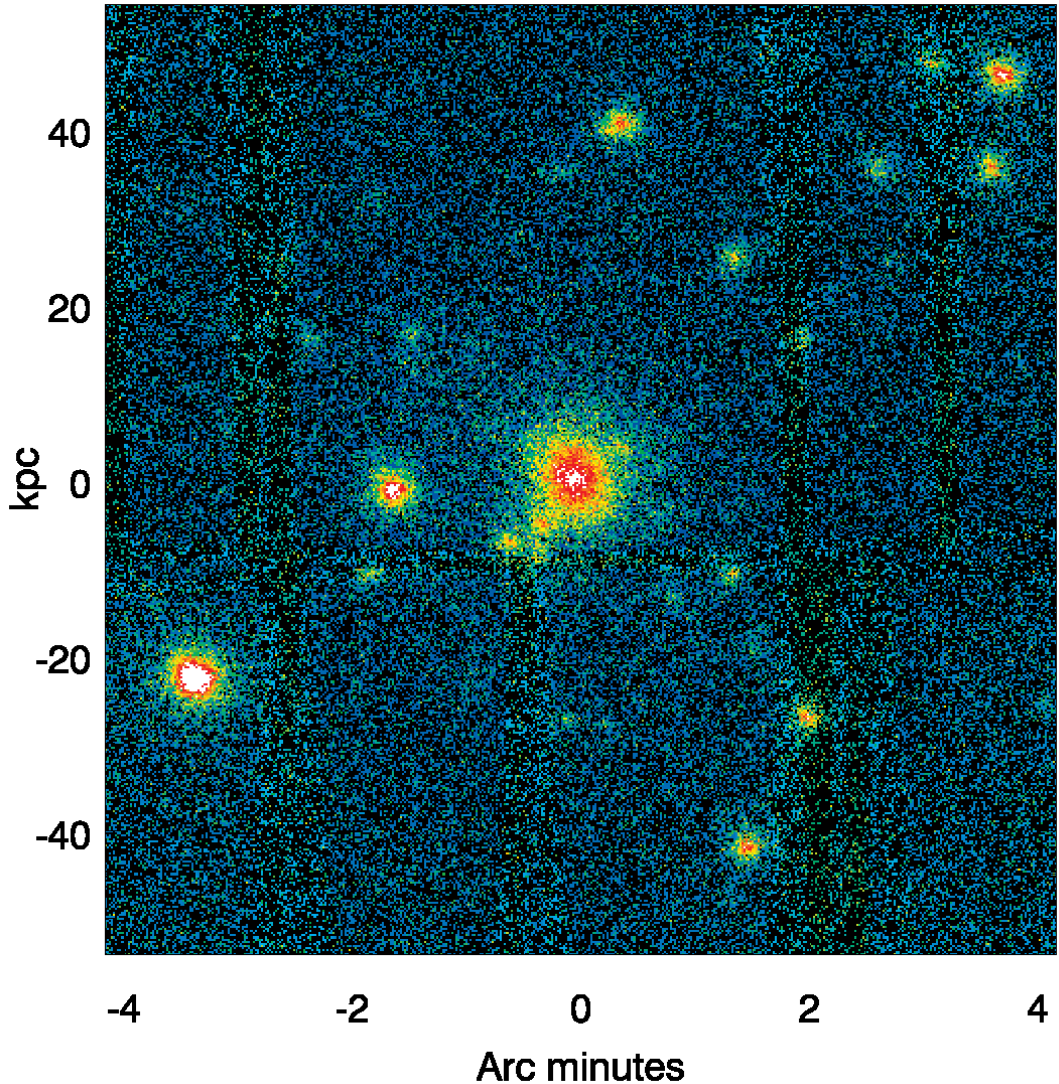}
\vspace{0.2cm}
      \caption{$0.3-2.5$ keV band \textit{XMM-Newton} image of NGC~6753. To create this image, we combined the data from all three observations and from the PN and two MOS cameras. The presence of extended X-ray emission associated with NGC~6753 is apparent, but it cannot be traced beyond the innermost $\sim1\arcmin$ region due to the  relatively high background level and the low signal-to-noise ratio. In addition, a notable population of luminous point sources are present, most of which originate from background AGN or X-ray binaries. The emission from these sources significantly contaminates the extended emission, hence we masked them from the analysis of the diffuse emission.}
     \label{fig:image}
  \end{center}
\end{figure*}

To build a more comprehensive picture about the state of the hot gas in the dark matter halos of massive spiral galaxies, deep X-ray observations are required. Out of the four galaxies with detected coronae, NGC~266 likely has a relatively low, $kT \sim 0.2$ keV, gas temperature, which -- combined with its large distance -- does not make it an ideal target \citep{bogdan13b}. While UGC 12591 is a very massive galaxy, it is located at 100 Mpc, and an 80 ks XMM-Newton observation was not sufficient to measure the gas temperature beyond the optical extent of the galaxy \citep{dai12}. NGC~1961 was analyzed by \citet{anderson16}, who utilized a 200~ks \textit{XMM-Newton} observation. These authors revealed the temperature structure of the hot gas and built a metallicity profile representing the first detailed study of the thermal and metallicity structure of the gas. However, one caveat with NGC~1961 is that  it may have undergone a minor merger in the recent past. This implies that the dark matter halo of NGC~6753 may be disturbed and its hot gas content may be contaminated. 

To extend our understanding about the in-depth characteristics of gaseous coronae, we utilize deep $290$~ks  \textit{XMM-Newton} observations of the spiral galaxy, NGC~6753. This galaxy is an ideal candidate for the detailed study of its gaseous coronae. Indeed, out of the four detected coronae NGC~6753 is the most nearby system, it is not a member of a rich galaxy group or cluster, it is not a starburst galaxy ($\rm{SFR} = 15.5 \ \rm{M_{\odot} \ yr^{-1}} $), and has relatively low line-of-sight column density. In Figure \ref{fig:multi} we show the \textit{GALEX} near-ultraviolet, the \textit{DSS} R-band, and the \textit{2MASS} K-band images \citep{jarrett03}. These images point out the nearly face-on\footnote{HyperLeda: http://leda.univ-lyon1.fr/} orientation of the galaxy ($i=30\degr$). NGC~6753 is compact relative to its large stellar mass ($M_{\rm \star} = 3.2 \times10^{11} \ \rm{M_{\odot}}$) since the dominant fraction of the stellar light and the associated star-formation activity is confined to $r\lesssim1.2\arcmin$ ($\lesssim15$ kpc) radius. Moreover, these images reveal the relaxed nature of the galaxy, implying that its dark matter halo and the hot gas within this halo was not disturbed by a recent interaction or merger. In addition, NGC~6753 is an ideal candidate to probe the characteristics of its gaseous corona, as our previous X-ray observations pointed out that the average temperature of its  corona is hot ($kT\sim0.6$ keV) and it is  luminous \citep{bogdan13a}, hence readily observable with \textit{XMM-Newton}. 

The goal of this study is to characterize the thermal and metallicity structure of the hot corona and probe the baryon content of NGC~6753. These detailed observational measurements can be used as powerful constraints to probe the predictions provided by present and future galaxy formation models. 

For NGC~6753 we adopted a distance of $D = 43.6$ Mpc   at which distance $1\arcmin = 12.67$ kpc. Following \citet{bogdan13a}, we assume that the virial mass of the galaxy is $M_{200}\sim 1.0\times10^{13} \ \rm{M_{\odot}}$ and its virial radius is $r_{200}\sim440$ kpc.  The Galactic column density towards NGC~6753 is $  N_{\rm H} = 5.4 \times 10^{20} \ \rm{cm^{-2}}$ \citep{kalberla05}. Throughout the paper we assume $H_0=71 \ \rm{km \ s^{-1} \ Mpc^{-1}}$, $ \Omega_M=0.3$, and $\Omega_{\Lambda}=0.7$, and that all error bars are $1\sigma$ uncertainties. 

This paper is structured as follows. In Section 2 we present the X-ray data and discuss their analysis. In Section 3 we present the surface brightness, density, temperature, and metallicity profiles. In Section 4 we discuss our results and we summarize in Section 5.

\begin{figure*}[t]
  \begin{center}
    \leavevmode
      \epsfxsize=0.8\textwidth\epsfbox{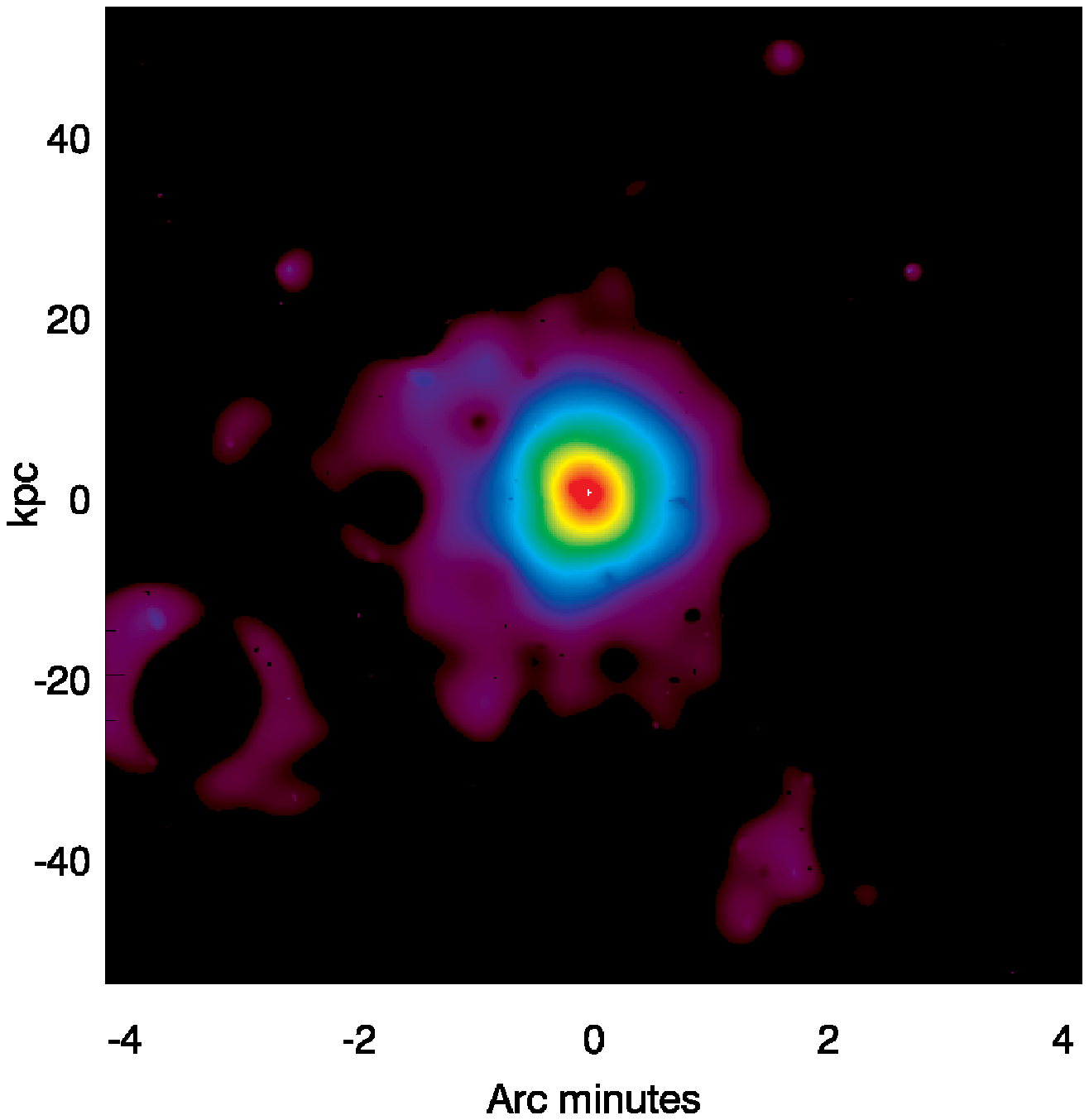}
\vspace{0.2cm}
      \caption{The $0.3-2.5$ keV band ``denoised'' image of the diffuse emission. This image combines the data from the three individual observations and three cameras aboard \textit{XMM-Newton}.  All of the background components are subtracted, the vignetting effects are corrected, and the resolved point sources are masked out. The image was reconstructed from wavelet coefficients thresholded at 3$\sigma$.  The image traces the diffuse X-ray emission out to $\sim2\arcmin$  ($\approx25$ kpc) from the galaxy center and reveals the symmetric nature of the gaseous emission.}
     \label{fig:image2}
  \end{center}
\end{figure*}

\begin{table}
\caption{List of analyzed \textit{XMM-Newton} observations.}
\begin{minipage}{8cm}
\renewcommand{\arraystretch}{1.3}
\centering
\begin{tabular}{c c c c c c}
\hline 
 \textit{XMM-Newton}  &$t_{\rm tot}$ & $ t_{\rm filt} $ &$t_{\rm filt} $  & $ t_{\rm filt} $ & Observation  \\ 
 Obs ID & & MOS1 &  MOS2 & PN & Date \\
& (ks) & (ks) & (ks) & (ks)  \\
\hline
0673170201 & 73.9 & 58.7 & 59.2 & 27.9 & Apr 21, 2012 \\
0782430101 & 108.0 & 87.5 & 87.1 & 50.0 &  Apr 12, 2016 \\
0782430201 & 108.0 & 93.1 & 94.3 & 36.6 & Apr 14, 2016  \\

\hline \\
\end{tabular} 
\end{minipage}

\label{tab:list2}
\end{table}

\newpage

\section{Data reduction}
\subsection{X-ray observations}
\label{sec:xmm}
NGC~6753 was observed with the European Photon Imaging Camera (EPIC) aboard XMM-Newton in three observations for a total of 290 ks. Details about the observations are listed in Table  \ref{tab:list2}. 

The data analysis, including the modeling of the instrumental and sky background components, is identical with that of our previous works \citep{bourdin08,bourdin13,bogdan13a}. For details, we refer to these earlier studies, and here we only summarize the main points of the analysis. The event lists were processed using the XMM Science Analysis System (SAS) version 16.0.0 and Current Calibration Files (CCF). Flare contaminated time intervals  were filtered by identifying high background periods in the light curves extracted from the $1-5$ keV and $10-12$ keV energy ranges. The exposure time of individual observations before and after flare filtering is listed in  Table \ref{tab:list2}. Overall, the EPIC PN  observations are significantly more affected by the high background periods than the EPIC MOS data. 

To optimize the signal-to-noise ratio of the imaging data, we combined the EPIC PN and EPIC MOS observations and merged the data from the three pointings. Luminous point sources could significantly contaminate the diffuse emission. Therefore, we identified these sources following the description of \citet{bogdan13a} and masked them from further analysis of the diffuse emission. Instrumental and sky background components were spatially and spectrally modeled. The amplitudes of the background components were jointly fitted within an annular region centered on the galaxy, in a galactocentric radius range of $75-115$ kpc, where the galaxy signal is expected to be negligible. As discussed in our previous studies, this method yields better than $\lesssim5\%$ background subtraction accuracy in the $0.3-2.5$ keV energy range.

\begin{figure*}[t]
  \begin{center}
    \leavevmode
      \epsfxsize=0.48\textwidth\epsfbox{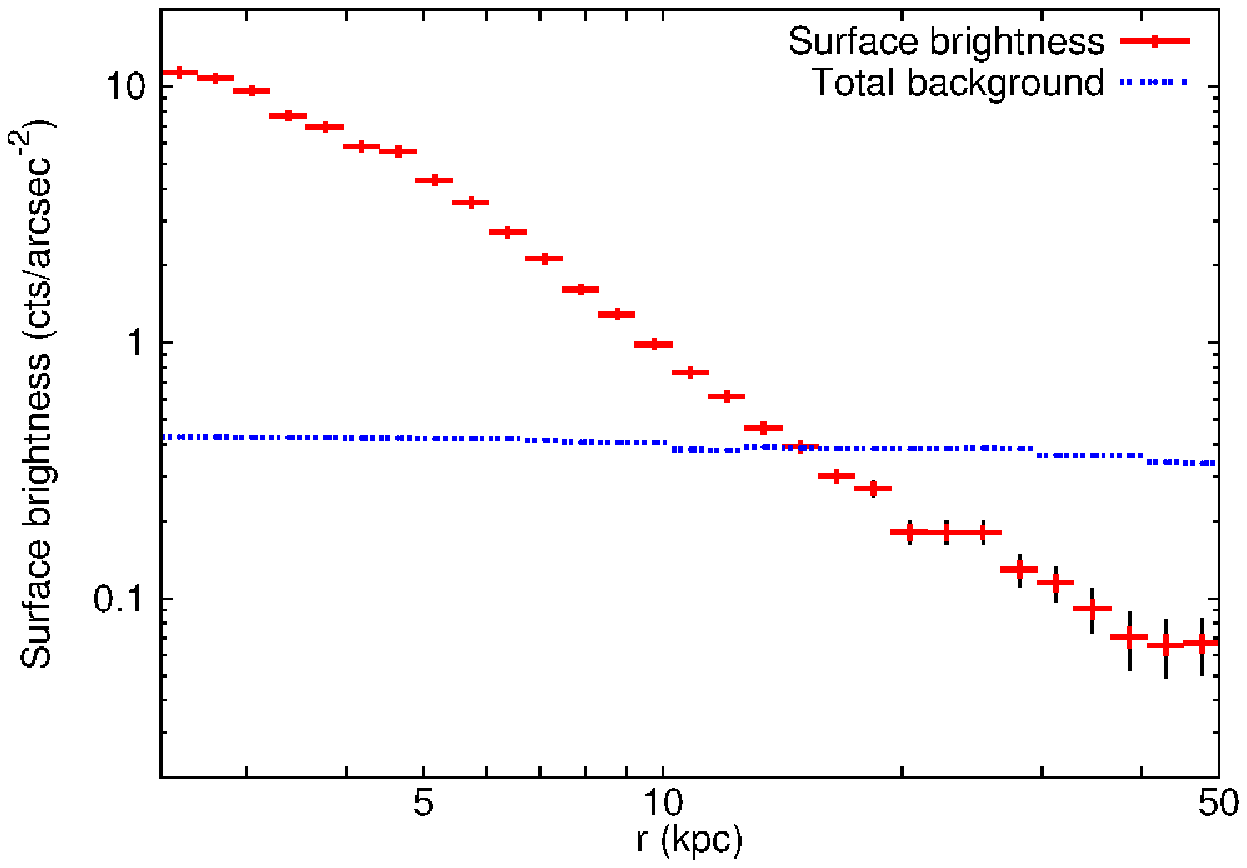}
      \epsfxsize=0.48\textwidth\epsfbox{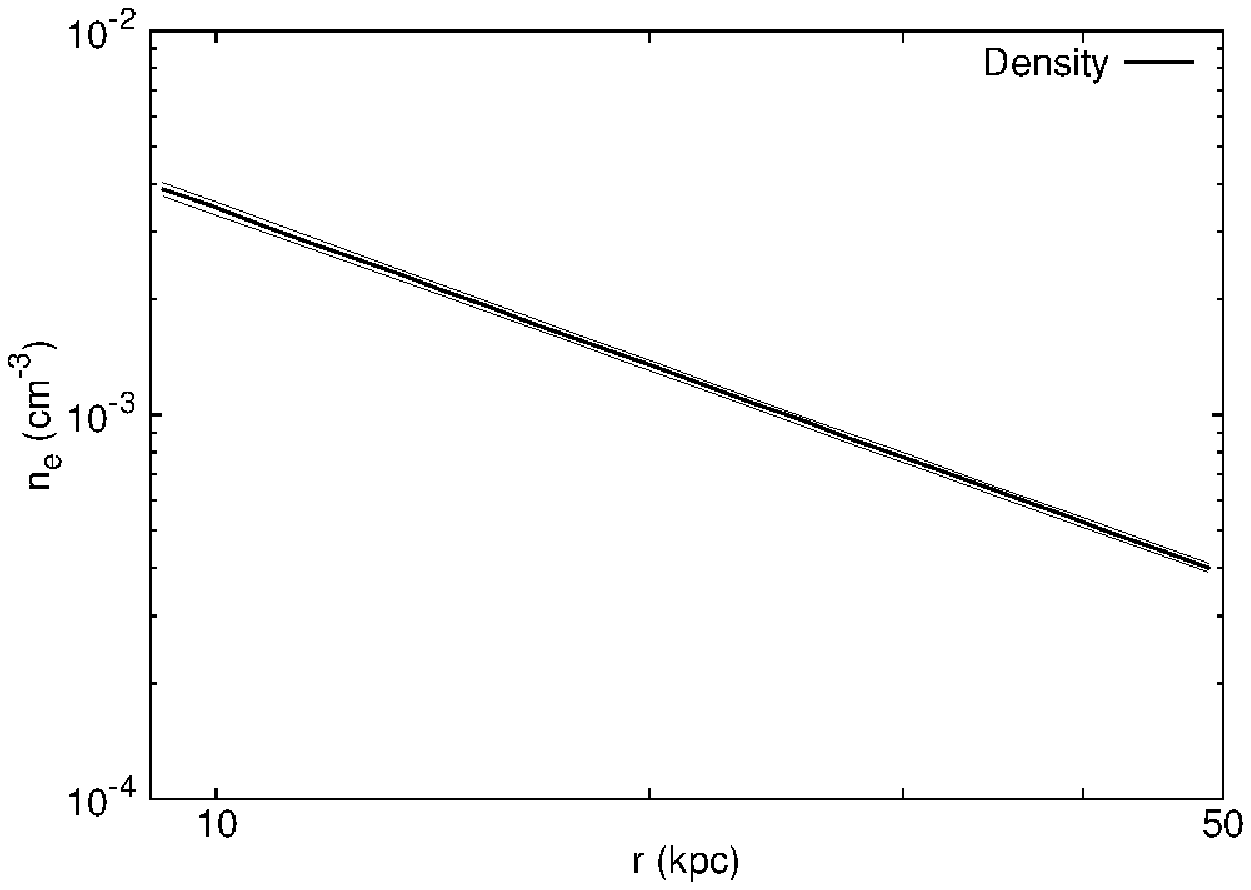}
\vspace{0.0cm}
      \caption{The left panel shows the surface brightness distribution of the extended X-ray emission in the $0.3-2.5$ keV energy range. The thick red error bars represent the statistical uncertainties, while the thin black error bars correspond to $5\%$ systematic uncertainty. The approximately horizontal line represents the combined background level that is already subtracted from the profile. The right panel shows the model of the electron density distribution of the hot gas. To obtain this model, the surface brightness distribution was fit using a modified $\beta$-model (see Section \ref{sec:sb}). The thin lines represent the uncertainties of the best-fit profile.}
     \label{fig:sbprof}
  \end{center}
\end{figure*}

\section{Results}
\subsection{Images}
\label{sec:images}

In the left panel of Figure \ref{fig:image} we show the combined \textit{XMM-Newton} image of NGC~6753 in the $0.3-2.5$ keV band. This image is corrected for spatial variations of the effective area and background. It reveals the presence of luminous X-ray sources in the vicinity of the galaxy, which were excluded from the analysis of the diffuse emission (see Section \ref{sec:xmm}).  The typical size of the excluded source regions was in the range of $10\arcsec-30\arcsec$ depending on the luminosity of the point sources. In addition, prominent diffuse X-ray emission is associated with the central regions of the galaxy, which has two different origins. Part of this emission is associated with the population of unresolved faint X-ray binaries, which is dominated by low-mass and high-mass X-ray binaries \citep{grimm03,gilfanov04,mineo12}, with other sources -- such as accreting white dwarfs, coronally active binaries, young stellar objects and young stars -- also contributing to a lesser significance \citep{sazonov06,revnivtsev06,bogdan08,bogdan11}. Since these discrete X-ray emitting components follow either the stellar light distribution or the star-formation activity, they are confined within the optical body of the galaxy. Additionally, part of the diffuse emission originates from truly diffuse gaseous emission. However, within the central regions of the galaxy the X-ray emitting gas may not only arise from the infall of primordial gas, but it is likely that at least part of the gas originates from stellar yields that are heated to X-ray temperatures by the energy input from supernovae \citep[e.g.][]{david06,li07,bogdan08}. Given that the gaseous components with different origins could mix and may be heated to similar temperatures, they are virtually indistinguishable. Therefore, it is more desirable to explore the hot corona beyond the stellar body of the galaxy. Indeed, beyond the optical radius the main X-ray emitting component is  the hot gas that likely originates from the infall of primordial gas. 

Although the hot gas can be traced out to 50 kpc from the center of NGC~6753 (see Section \ref{sec:sb} for details), due to the low signal-to-noise ratio the presence of the large-scale diffuse emission is not evident in the raw image. To better explore the diffuse gaseous emission, we depict the denoised surface brightness image of NGC~6753 in Figure \ref{fig:image2}. This image is the result of a B3-spline wavelet filtering using four spatial scales that sample details in the range of $1.5\arcmin-13\arcmin$ arcmin. In this procedure, the X-ray surface brightness is corrected for spatial variations of the effective area and background, and the noise fluctuation is approximated as a Gaussian process by using a multiscale variance stabilizing transformation \citep{
starck09}. This image traces the diffuse emission out to $\sim2\arcmin$ radius and demonstrates the isotropic gas distribution around NGC~6753, hinting that the hot gas likely resides in hydrostatic equilibrium. 

Although the images demonstrate the presence of X-ray emission extending beyond the optical radius of the galaxy, to perform a quantitative analysis of the gas we study its surface brightness and temperature structure in the next sections.

\begin{table}
\caption{Best-fit parameters of the density profile.}
\renewcommand{\arraystretch}{1.5}
\centering
\begin{tabular}{c c c c c c c c}
\hline 
Galaxy&  $n_0$ &$r_c $ & $r_s^{\dagger}$ & $\alpha$ &$\beta$$^{\dagger}$  & $\epsilon^{\dagger}$  & $\gamma^{\dagger}$ \\ 
 & ($ \rm{cm^{-3}}$) & (kpc)  & (kpc) & &   \\ 
\hline
NGC~6753 & $1.8\times10^{-3}$ & 3.90 & 8.26 & 2.73 & 0.30& 0 & 3.0 \\
\hline \\
\end{tabular} 
$^{\dagger}$ These parameters were fixed at the given value.
\label{tab:fit}
\end{table}

\begin{figure}[t]
  \begin{center}
    \leavevmode
      \epsfxsize=0.48\textwidth\epsfbox{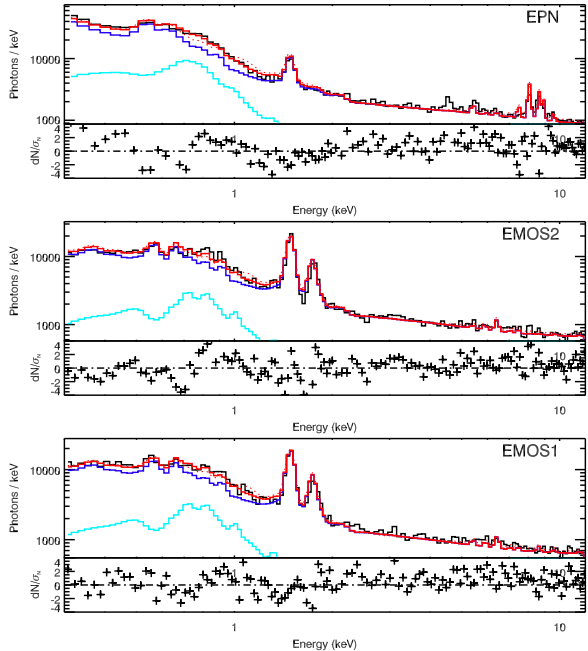}
	\vspace{0.2cm}
      \caption{X-ray energy spectrum of NGC~6753 extracted from the $20$ to $50$ kpc radial region. The top, middle, and bottom panels show the spectra extracted from the EPIC PN, EPIC MOS1, and EPIC MOS2 cameras. The spectra are fit with an absorbed optically thin thermal plasma emission model. The data and the overall best-fit model are shown with the black and red lines, respectively. The combined emission from the background is shown with the dark blue lines, and the emission associated with the thermal spectrum is shown with the light blue lines.}
\vspace{0cm}
     \label{fig:spectra}
  \end{center}
\end{figure}

\begin{figure*}[t]
  \begin{center}
    \leavevmode
      \epsfxsize=0.48\textwidth\epsfbox{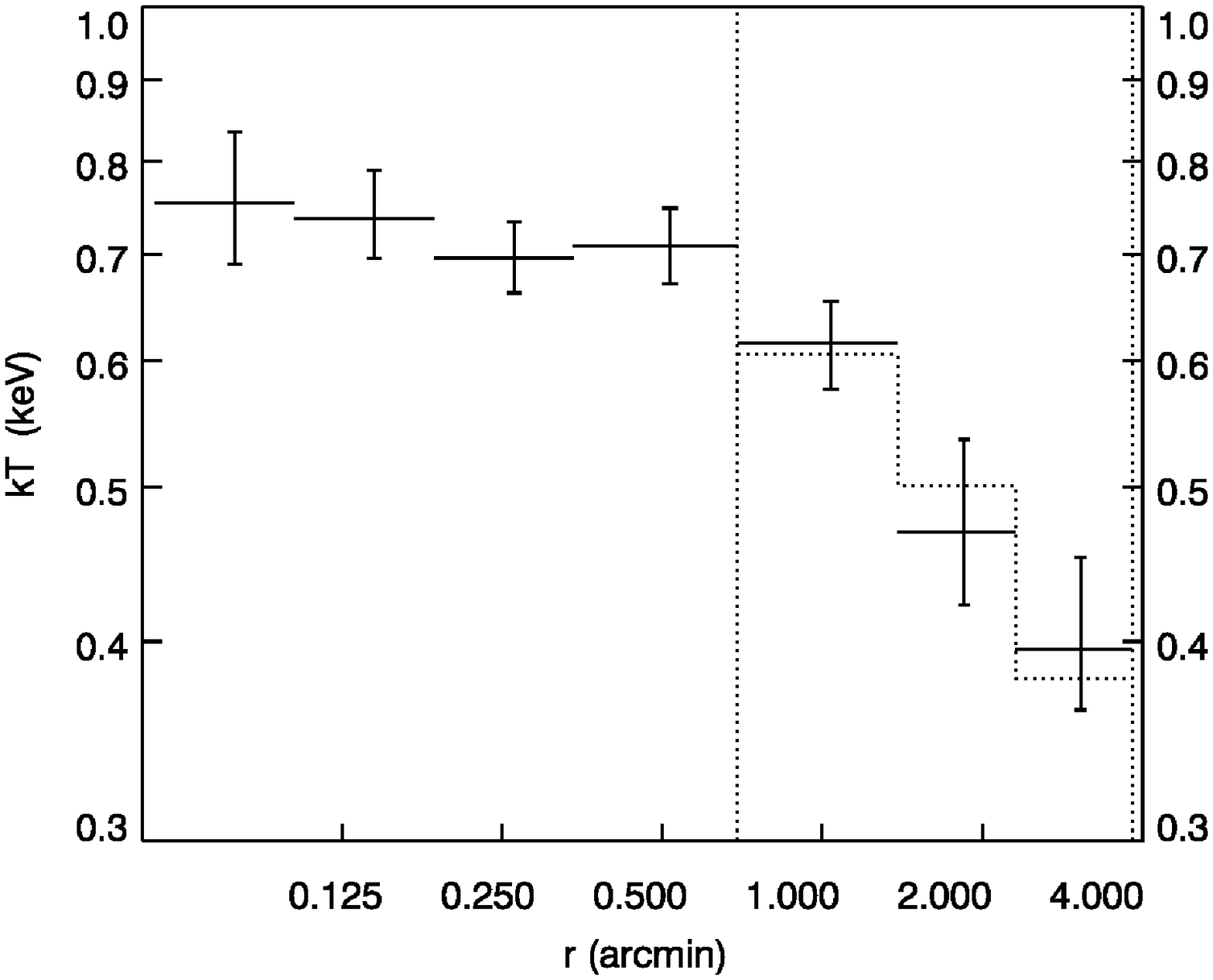}
      \epsfxsize=0.48\textwidth\epsfbox{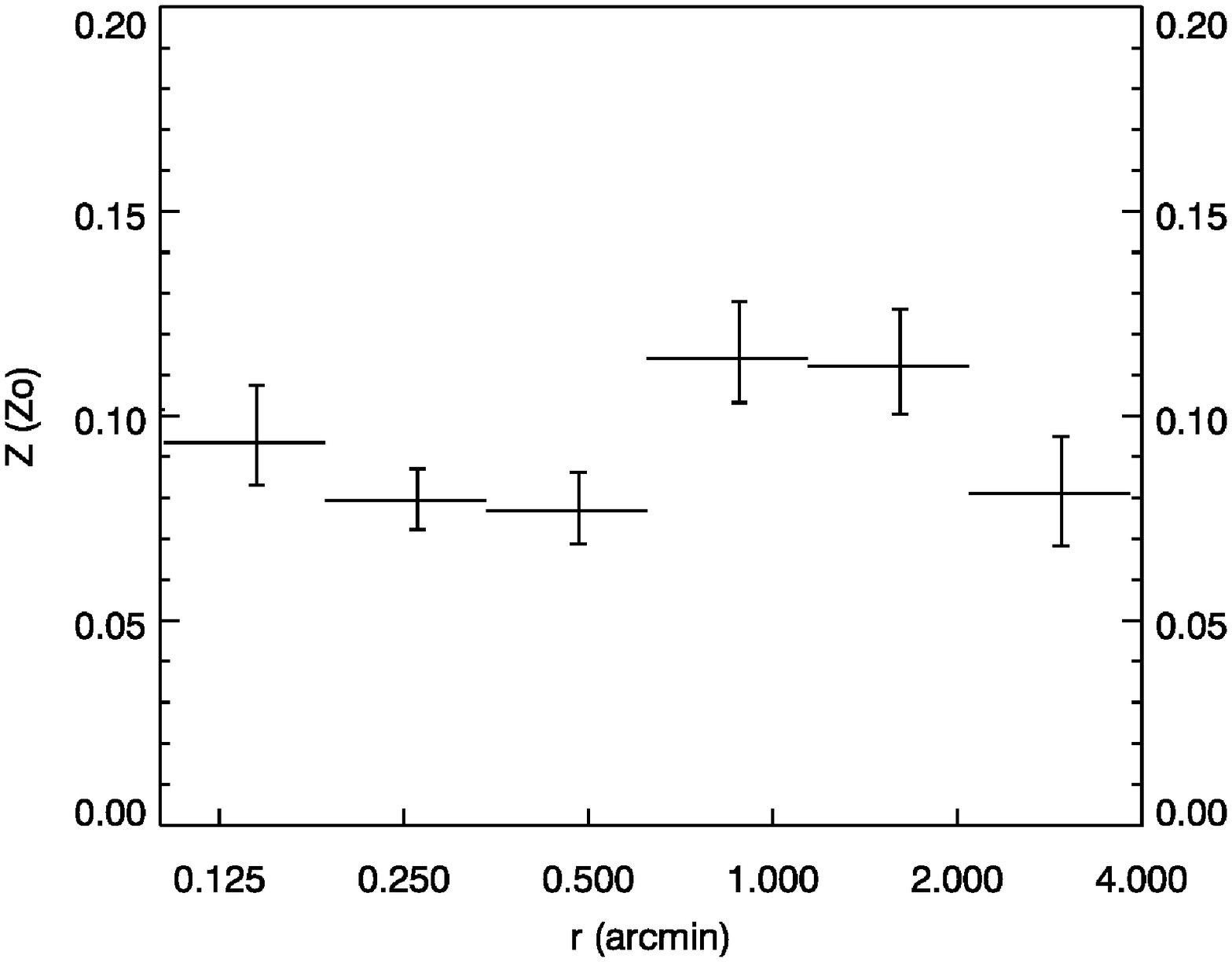}
	\vspace{-0.5cm}
      \caption{Azimuthally averaged radial profile of the hot gas temperature (left panel) and metallicity (right panel). To obtain the best-fit values we described the thermal emission with a single temperature, optically thin thermal plasma emission model. During the fitting procedure both the temperature and metallicity were fit as free parameters. The dotted histogram shows the best-fit analytic temperature model. The temperature profile notes that the temperature of the hot gas decreases with an increasing radius and drops to $\sim0.4$ keV beyond $2\arcmin$. However, the metallicity is virtually uniform at every radius at $Z\sim0.1$ Solar.}
     \label{fig:temp_abund}
  \end{center}
\end{figure*}

\subsection{Surface brightness and density profiles}
\label{sec:sb}

To probe the spatial distribution of the diffuse emission, we extract its surface brightness profile using circular annuli centered on the nucleus of NGC~6753. The $0.3-2.5$ keV band  surface brightness profile, shown in the left panel of Figure \ref{fig:sbprof}, was obtained by combining data from all three observations and cameras. To construct the profile, all point sources were excluded. All background components are subtracted and their level is shown with the approximately horizontal line.

The surface brightness profile robustly traces the gas out to  $\sim50$ kpc. Beyond this radius the signal from the hot gas drops rapidly and the signal-to-noise ratio does not exceed a factor 2, given the systematic uncertainties associated with the background subtraction. While statistical uncertainties would allow for the tracing of the hot gas beyond these radii, due to the systematics these results may be inaccurate. Therefore, we conservatively restrict our study to the hot gas that is confined within $50$ kpc of NGC~6753. 

Since the predominant fraction of the stellar light of NGC~6753 is confined to within $\approx15$ kpc \citep[see Figure \ref{fig:multi} and][]{bogdan13a}, it is clear from Figure  \ref{fig:sbprof} that the diffuse X-ray emission has a significantly broader distribution. As discussed in Section \ref{sec:images}, at radii $\gtrsim15$ kpc  the  diffuse emission is dominated by truly diffuse gas and the emission from unresolved compact objects associated with the galaxy does not play a significant role. In addition, the hot gas at these larger radii is likely dominated by the accreted primordial gas and is not recycled gas that was expelled from evolved stars and heated to X-ray temperatures. Taking these together, it is the most ideal to study the diffuse gaseous emission between the 20-50 kpc radial range. 

From the surface brightness distribution of the hot gas we derive its density profile. We fit the surface brightness profile in the $0.5\arcmin-4\arcmin$  ($6.3-50.7$ kpc) region with a modified $\beta$-model described by \citet{vikhlinin06}. By assuming a constant metallicity (see Section \ref{sec:metallicity}), we compute the emission measure profile as a function of the radius using

\begin{eqnarray}
 n_{p} n_{e}=n_0^2 \frac{(r/r_c)^{-\alpha}}{(1+r^2/r_c^2)^{3\beta-\alpha/2}} \frac{1}{(1+r^\gamma/r_s^{\gamma})^{\epsilon/\gamma}} \ ,
\end{eqnarray}

The main advantage of this more complex model over a traditional $\beta$-model is that it can accurately fit a wide range of X-ray surface brightness profiles. For details on the individual parameters we refer to \citet{vikhlinin06}. The best-fit parameters of the  density profile are tabulated in Table \ref{tab:fit}. The density profile is depicted in the right panel of Figure \ref{fig:sbprof}.  The profile shows that the electron density drops from $1.3\times10^{-3} \ \rm{cm^{-3}}$ at $20$ kpc to $0.4\times10^{-3} \ \rm{cm^{-3}}$ at $50$ kpc.

\begin{table}
\caption{Best-fit parameters of the spectral fits.}
\renewcommand{\arraystretch}{1.5}
\centering
\begin{tabular}{c c c}
\hline 
Region& $kT$ & $Z_{\rm \odot}$\\ 
 (arcmin)& (keV) & (Solar)  \\ 
\hline
$0.055-0.102$ & $0.75^{+0.08}_{-0.06}$ & $0.102^{+0.021}_{-0.016}$  \\
$0.102-0.187$ & $0.74^{+0.05}_{-0.04}$ & $0.094^{+0.012}_{-0.010}$  \\
$0.187-0.344$ & $0.70^{+0.04}_{-0.04}$ & $0.079^{+0.007}_{-0.007}$ \\
$0.344-0.694$ & $0.71^{+0.04}_{-0.03}$ & $0.077^{+0.009}_{-0.007}$  \\
$0.694-1.381$ & $0.61^{+0.04}_{-0.04}$ & $0.118^{+0.013}_{-0.011}$  \\
$1.381-2.314$ & $0.47^{+0.08}_{-0.05}$ & $0.124^{+0.019}_{-0.016}$  \\
$2.314- 3.837$ & $0.39^{+0.09}_{-0.03}$ & $0.075^{+0.014}_{-0.012}$ \\
\hline \\
\end{tabular} 
\label{tab:spectra}
\end{table}

\begin{figure*}[t]
  \begin{center}
    \leavevmode
      \epsfxsize=0.8\textwidth\epsfbox{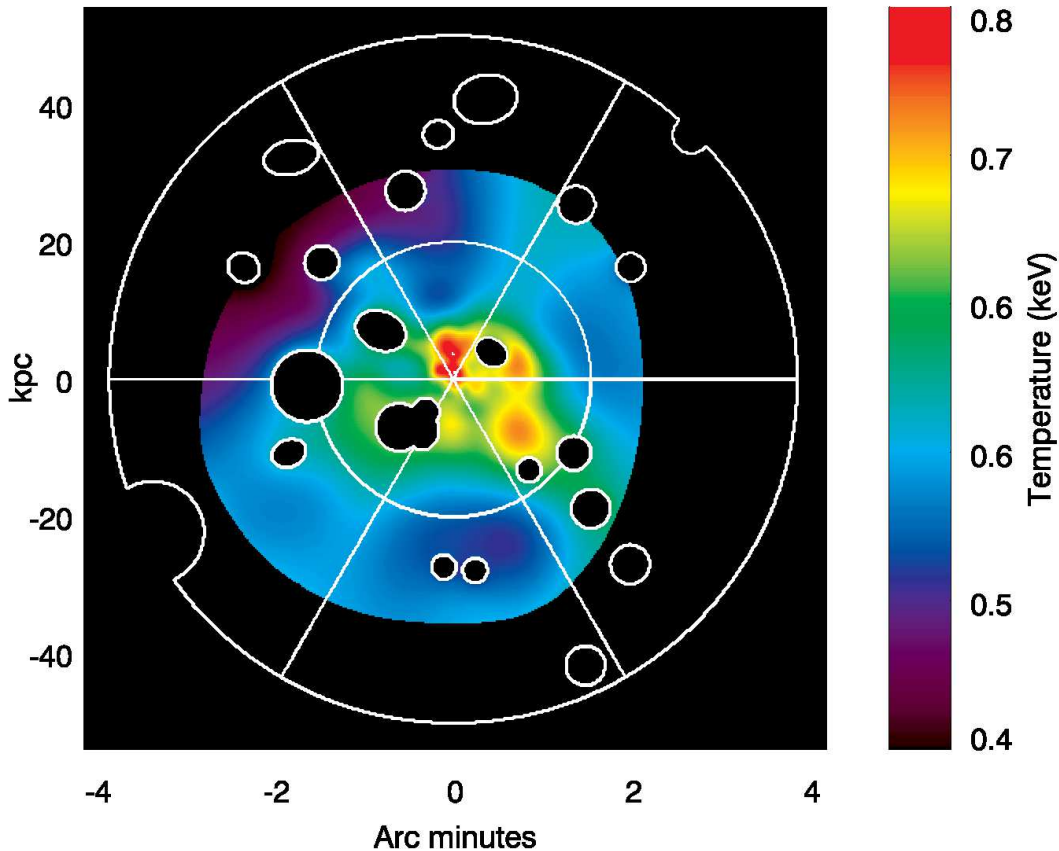}
\vspace{0.2cm}
      \caption{Temperature map of the hot gaseous emission around NGC~6753. To construct the temperature map, we applied B-spline wavelet filtering following \citet{bourdin08}. The angular sectors ($0-20$ kpc and $20-50$ kpc annuli) that were used to extract the temperature profiles depicted in Figure \ref{fig:wedges} are over plotted. Due to the low signal-to-noise ratios, an accurate temperature map cannot be constructed beyond $\sim30$ kpc radius. Note that the temperature map reveals the complex temperature structure of the hot gas with hotspots and cooler regions.}
     \label{fig:tmap}
  \end{center}
\end{figure*}

\begin{figure*}[t]
  \begin{center}
    \leavevmode
      \epsfxsize=0.48\textwidth\epsfbox{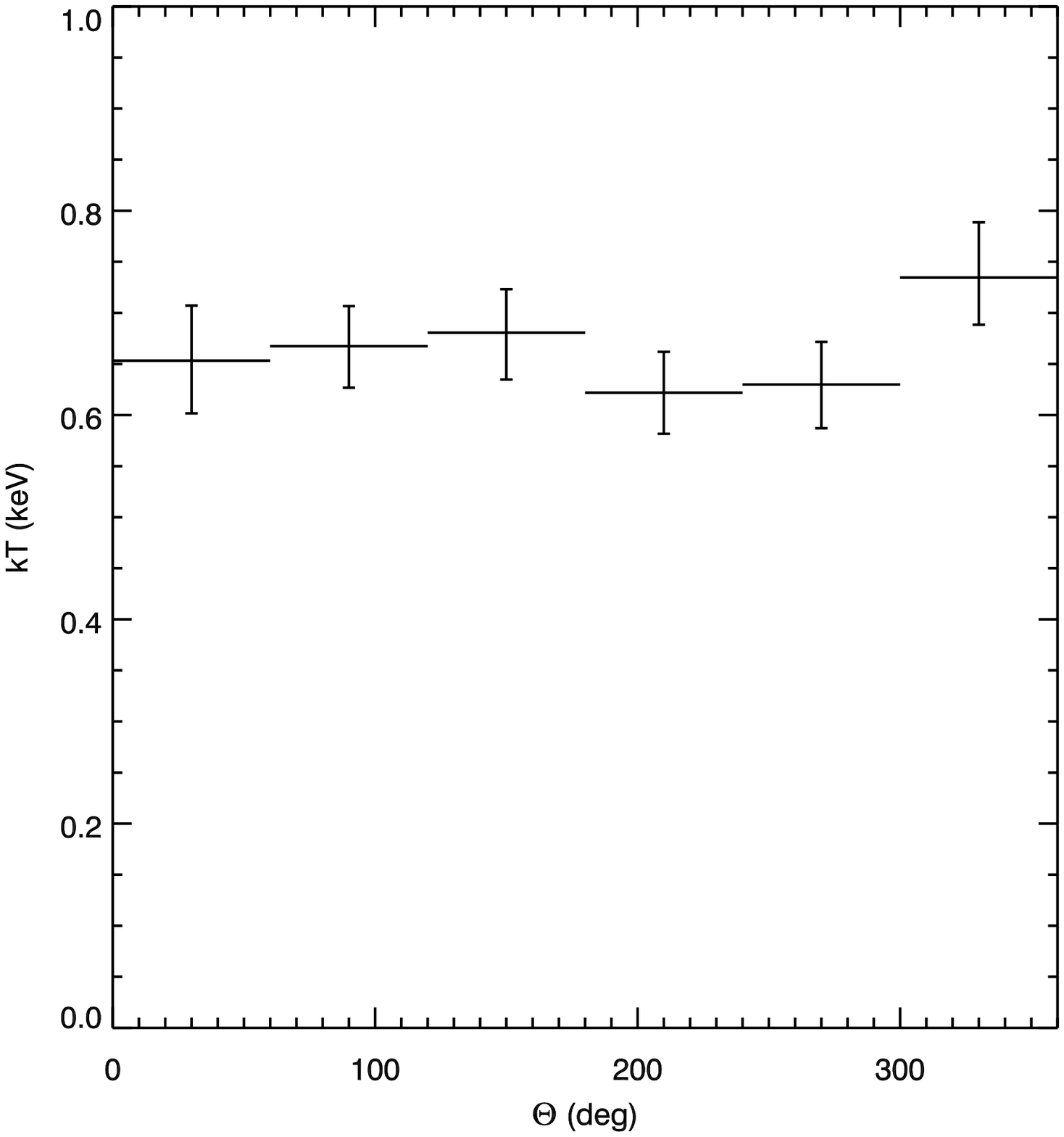}
      \epsfxsize=0.48\textwidth\epsfbox{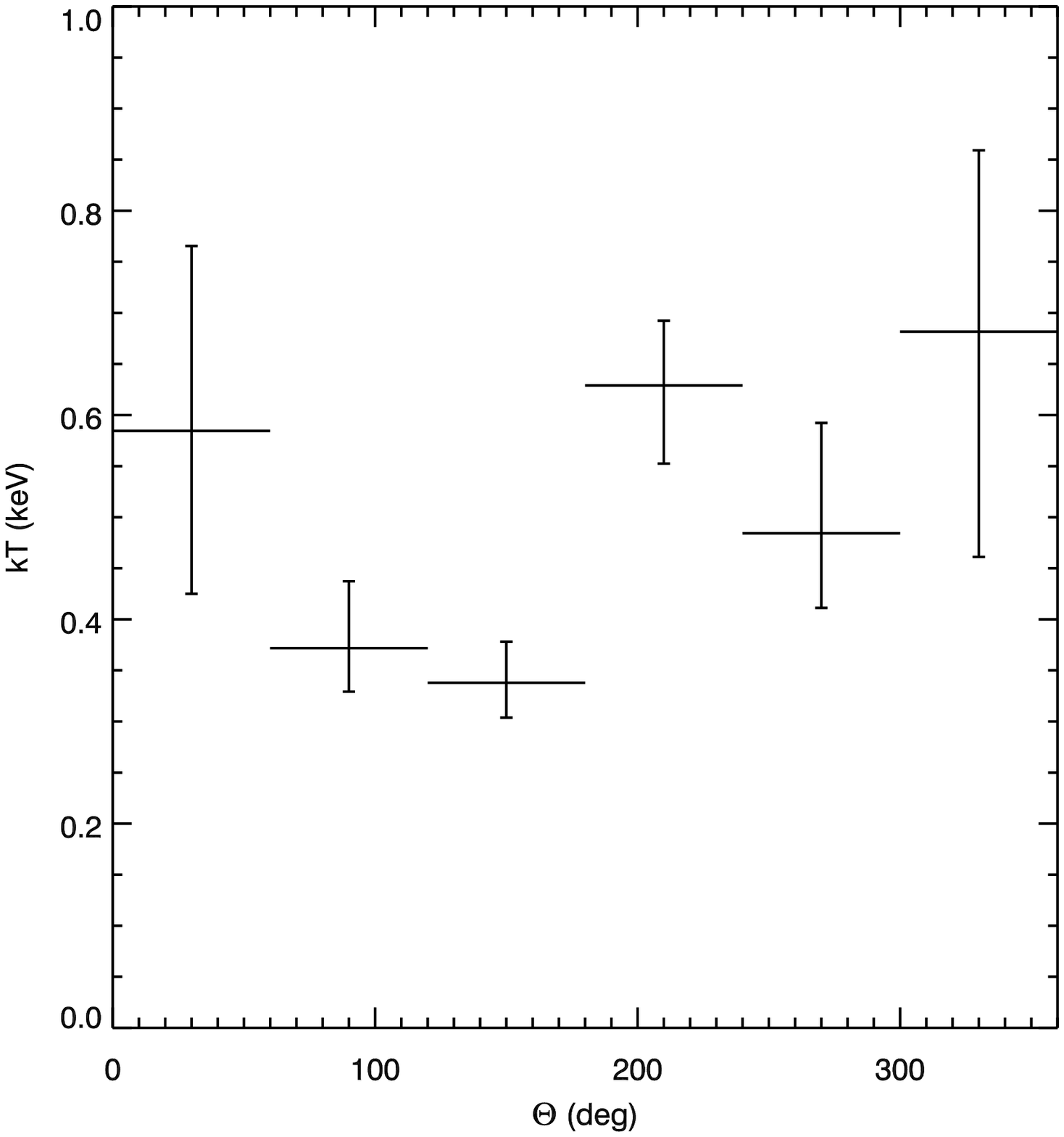}      
\vspace{0.2cm}
      \caption{Temperature distribution of the hot gas measured in circular wedges in the $0-20$ kpc (left panel) and $20-50$ kpc (right panel) regions. The position angles $0\degr$ and $90\degr$ correspond to east and north, respectively. The depicted values are the best-fit temperatures obtained from the spectra extracted from each wedge. Despite the small scale temperature fluctuations (see Figure \ref{fig:tmap}),  the average temperature within 20 kpc is uniform at $kT\approx0.6$ keV. However, the profile extracted from the $20-50$ kpc region indicates notable temperature structures and hints that the gas is cooler in the $60\degr-180\degr$ region than anywhere else.}
     \label{fig:wedges}
  \end{center}
\end{figure*}

\subsection{Temperature and metallicity profiles}
\label{sec:metallicity}

Thanks to our deep $290$ ks \textit{XMM-Newton} observations, we are in a unique position to probe the temperature and metallicity structure of the gaseous corona around NGC~6753 in unprecedented detail. To this end, we extract X-ray energy spectra of the diffuse emission around NGC~6753. Given the different response functions of the three cameras, we extract individual spectra for the PN and two MOS cameras and fit the spectra simultaneously. 

To fit the hot gaseous emission, we utilize an optically thin thermal plasma emission model  (\textsc{apec} in \textsc{Xspec}), which provides measurements of the temperature and metallicity of the gas. Therefore, both of these parameters were allowed to vary during the fitting process. We note that we used the abundance table from \citet{grevesse98}. For all regions we also added an absorption component that was fixed at the Galactic value ($N_{\rm H} = 5.4 \times 10^{20} \ \rm{cm^{-2}}$). In Figure \ref{fig:spectra} we show the representative X-ray energy spectrum of the $20-50$ kpc radial region, which shows the contribution of background components, the emission originating from the hot gas, and the overall quality of the fit. 

To study the azimuthally averaged radial profiles of the gas temperature and metallicity, we extracted X-ray spectra of seven regions using circular annuli centered on the nucleus of NGC~6753. Using the above described fitting procedure, we construct temperature and metallicity profiles that are depicted in the left and right panels of Figure \ref{fig:temp_abund}, respectively. The best-fit values are listed in Table \ref{tab:spectra}. 

The radial temperature profile reveals that the gas temperature is non-uniform and shows a decreasing trend toward larger radii. Indeed, while within the optical body of the galaxy we measure  $kT\approx0.75$ keV, the temperature drops to $kT\approx0.4$ keV beyond $2\arcmin$. The observed temperature profile is similar to that obtained for NGC~1961  \citep{anderson16} and suggests that the decrease in gas temperature toward larger radii may be a common phenomena in massive isolated disk galaxies. We note that the observed temperature gradient is similar to that observed in X-ray faint elliptical galaxies, which show flat or declining temperature profiles \citep[e.g.][]{fukazawa06,humphrey06}. These galaxies, similarly to NGC~6753 and NGC~1961, are preferentially located in relatively isolated environments. Therefore, the luminous X-ray atmospheres of rich galaxy groups or clusters do not dominate over the observed emission from galaxy coronae at larger radii. Alternatively, it is also possible that the negative temperature gradients are due to the presence of partial winds \citep[e.g.][]{pellegrini98,mathews03}, in which scenario the outflow is suppressed in the central regions of the galaxy, but a (subsonic) wind is maintained at the outer parts of galaxies.

Although, the temperature profile suggests an overall decrease as a function of an increasing radius, based on this profile it cannot be established whether the hot gas exhibits a notable temperature structure.  To explore the presence of any temperature structure around the gas, we use two approaches. First, we construct a denoised  temperature map of NGC~6753. The image, depicted in Figure \ref{fig:tmap}, was derived following the spectral-imaging algorithm described in \citet{bourdin08}. This temperature map explores the temperature structure of the gas within $30$ kpc radius, since at larger radii the signal-to-noise ratio is not sufficiently high enough to construct the map. This temperature map points out the complex temperature structure of the gas within the inner $30$ kpc region. Specifically, we detect a hotspot in the north and a hotter filamentary region on the southwest of the galaxy. Second, we extracted energy spectra using circular wedges in the $20-50$ kpc region with opening angles of $60\degr$. The spectra were fit with an optically thin thermal plasma emission model. In this temperature profile, shown in  Figure \ref{fig:wedges},   $0\degr$ and $90\degr$ correspond to east and north, respectively. Although the best-fit temperatures have notable uncertainties, this plot hints at the existence of temperature non-uniformities at the $1\sigma-2\sigma$ level. Specifically, between the position angles of $60\degr-180\degr$ we observe cooler, $kT\sim0.4$ keV, gas temperatures, while in other regions hotter ($kT\sim0.6$ keV) gas dominates. As a caveat, we mention that these temperature measurements are luminosity-weighted, hence the best-fit temperatures are more strongly influenced by the more luminous inner regions than those of the outer regions  of NGC~6753. Nevertheless, this temperature profile in in agreement with the temperature map and notes the complex temperature structure of the gas. 

The radial metallicity profile, presented in the right panel of Figure \ref{fig:temp_abund}, demonstrates that the hot gas has virtually uniform metallicity, $Z\sim0.1$ Solar, from the central regions out to $4\arcmin$ radius. To probe whether the observed spectra can be fit with higher metallicities, we fixed the metallicities at $0.3$ Solar and $1$ Solar and re-fit the spectra. The best-fit spectra, particularly the EPIC-PN data that has a larger collecting area, showed large residuals with higher metallicities, and the resulting reduced $\chi^2/\rm{d.o.f.}>2$ values were formally not acceptable. This conclusion holds whether we investigate narrow annuli (see the regions in Table \ref{tab:spectra}) or broader regions (e.g. an annulus with $20-50$ kpc radii). This implies that the low observed metallicities are a genuine characteristics of the hot gas in an around NGC~6753.

\subsection{Mass profiles}
\label{sec:hydro}

The presence of extended hot gaseous emission around NGC~6753, combined with the well-constrained density and temperature profile within 50 kpc radius, allows us to compute the total gravitating mass of the galaxy. Clearly, this measurement is independent from that obtained from other methods, such as that based on the circular velocity of the galaxy \citep{bogdan13a}. 

Assuming that the hot gas has a spherically symmetric distribution and resides in hydrostatic equilibrium in the gravitational potential well of the galaxy, we can infer the total gravitating mass using
 $$  M_{\rm{tot}} (<r) = - \frac{kT_{\rm{gas}}(r) r }{G \mu m_{\rm{p}}}  \Bigg( \frac{\partial \ln n_{\rm{e}}}{\partial \ln r} + \frac{\partial \ln T_{\rm{gas}}}{\partial \ln r} \Bigg) ,   $$
where $T_{\rm{gas}}$ and $n_{\rm{e}}$ are the deprojected temperature and density, respectively. 

We carry out the deprojection by using analytical distributions of $n_{\rm e}$ and $T_{\rm gas}$, where $n_{\rm e}$  is defined in Equation (1) and $T_{\rm gas}$ is a power law of the galactocenric radius. These 3D distributions are integrated along the line-of-sight and simultaneously fit to the X-ray surface brightness profile and the spectroscopic temperature profile. This procedure assumes a 2D convolution of the surface brightness with the \textit{XMM-Newton} point spread function, and a temperature weighting scheme first proposed in \citet{mazzotta04}, that mimics the spectroscopic response of a single temperature fit inside each temperature bin. To further derive confidence envelopes around $n_{\rm e}$, $T_{\rm gas}$, and $M_{\rm tot}$, this procedure is repeated on 400 random realizations of the surface brightness and on temperature profiles.  

\begin{figure}[t]
  \begin{center}
    \leavevmode
      \epsfxsize=0.48\textwidth\epsfbox{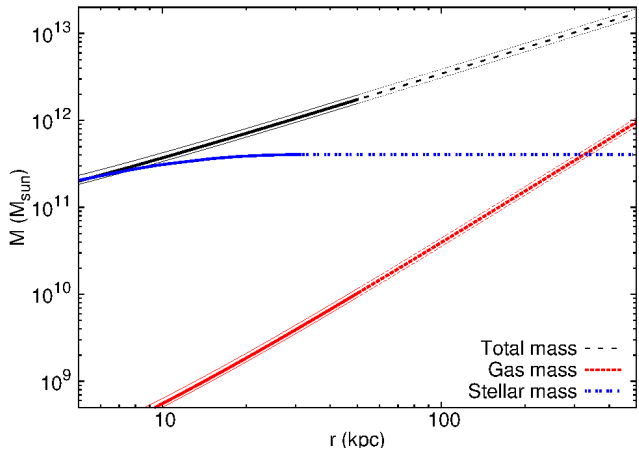}
       \vspace{-0.2cm}
     \caption{Mass profiles of NGC~6753. The black curves represent the total gravitating mass computed under the assumption of hydrostatic equilibrium (see Section \ref{sec:hydro}). The red curves depict the distribution of the hot gas mass. The blue curve shows the stellar mass distribution of NGC~6753. The total gravitating mass and the gas mass is measured within 50 kpc (solid lines), beyond which we extrapolate the distribution to the virial radius (dashed lines). The thin lines above and below the total mass and gas mass represent the statistical uncertainties.}
\vspace{0.5cm}
     \label{fig:mass}
  \end{center}
\end{figure}

The best-fit gravitating mass profile is shown in Figure \ref{fig:mass}, where we also depict the mass profiles of the hot gas and the stellar mass.  Since the density and temperature profiles are measured out to $50$ kpc, we must extrapolate beyond this radius. To this end, we extrapolate  the density profile by using the the best-fit model described in Section \ref{sec:sb} and the parameters listed in Table \ref{tab:fit}. For the temperature profile we assumed that beyond 50 kpc it remains constant $kT=0.4$ keV out to the virial radius. This choice is motivated by the observed temperature profile of galaxy groups and clusters that exhibit an approximately flat temperature profile at large radii \citep{lovisari15}. We estimate that the total mass within 440 kpc is $1.4\times10^{13} \ \rm{M_{\odot}}$, which is broadly consistent with that estimated based on the rotation curve of NGC~6753 \citep{bogdan13a}.  We emphasize that this value should be considered as an estimate since the density and temperature profiles are only traced out to $50$ kpc, which corresponds to $\approx11\%$ of the virial radius.

\section{Discussion}
\subsection{Baryon mass fraction}
\label{sec:baryon}

Measuring the baryon mass fraction of galaxies has important implications for galaxy evolution. Specifically, by probing the baryon content of galaxies, we can constrain the effects of AGN feedback and probe whether a notable fraction of gas was expelled from the gravitational potential well or was driven to larger radii. In galaxies, it is believed that, besides the stellar mass, the X-ray emitting gaseous component is the major constituent of the baryonic matter. Indeed, based on results from the COS-Halos survey it was established that only a fraction of the missing baryons are in the phase of cool ($T\sim10^4$ K) ionized gas \citep[e.g.][]{werk14}. Therefore, it is essential to constrain the total hot gas mass within the virial radius and constrain the baryon mass fraction of the galaxies. 

Given that our X-ray measurement can constrain the gas properties out to 50 kpc or $\sim11\%$ of the virial radius, we must extrapolate the gas mass profile out to the virial radius. To derive the total gas mass, we integrate the extrapolated density profile out to the virial radius and assume that the X-ray emitting gas has a constant $0.4$ keV temperature beyond 50 kpc. This extrapolation results in a total gas mass of  $4.0\times10^{11}  \ \rm{M_{\odot}}$. We emphasize that this extrapolation has notable systematic uncertainties, which have two main origins \citep[for details see][]{bogdan13a}. First, it stems from the fact that only a small fraction of the total volume is probed, and hence, about $97\%$ of the X-ray gas remains hidden from our X-ray study. Specifically, both the gas density and temperature profiles are extrapolated beyond 50 kpc radius. If these profiles deviate from the assumed profiles, that could significantly alter the estimated total gas mass. Second, if the metallicity of the gas differs from the best-fit value, that would result in a change in the emission measure and in the gas mass due to the degeneracy between the metallicity and the emission measure. Therefore, if the gas has notably higher metallicity than the observed $\sim0.1$ Solar value, the gas masses must be reduced.

\begin{figure}[t]
  \begin{center}
    \leavevmode
      \epsfxsize=0.48\textwidth\epsfbox{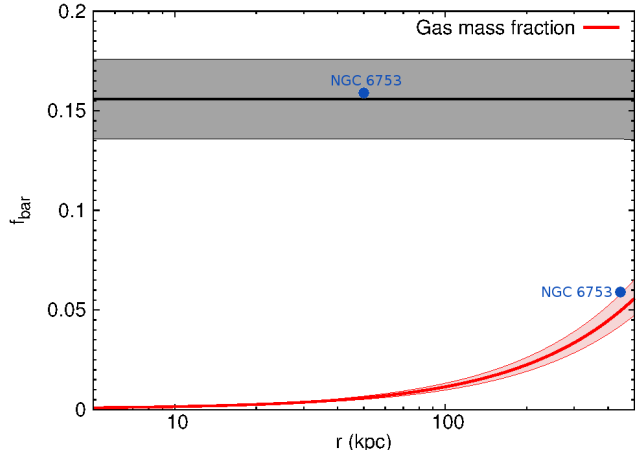}
\vspace{-0.2cm}
      \caption{Gas mass fraction as a function of radius for NGC~6753. Note that the curve only includes the hot gas mass and that other baryonic components are excluded. The shaded area represents the statistical uncertainties. The solid horizontal line and the shaded area correspond to the cosmic baryon mass fraction based on the findings of \citet{planck16}.  We show the total baryon mass fraction (including all baryonic components) of NGC~6753 within 50 kpc and within 440 kpc. Within 50 kpc the stellar mass dominates while the dark matter contributes to a lesser extent (see Figure \ref{fig:mass}), which results in $f_{\rm b} \approx 0.16$. Within 440 kpc the baryon mass fraction falls short of the cosmic value.} 
     \label{fig:fgas}
  \end{center}
\end{figure}

To compute the total baryon mass, we must include other baryonic components within the virial radius. Specifically, we must account for galaxies within the virial radius and gas in other phases. We derive the stellar mass from the \textit{2MASS} K-band images and by utilizing the K-band mass-to-light ratio ($M/L=0.81$) obtained from the relation established by \citet{bell03} and the $B-V$ color index.  We find that he total stellar mass of NGC~6753 is $M_{\rm \star}  = 3.2\times10^{11}  \ \rm{M_{\odot}}$. In addition, NGC~6753 holds a notable amount of HI gas, namely  $M_{\rm \star}  = 2\times10^{10}  \ \rm{M_{\odot}}$. Besides NGC~6753, several fainter galaxies are located in the proximity of NGC~6753, which have similar redshifts and are within the projected distance of the virial radius. Given the distance of $D=43.6$ Mpc of NGC~6753, we filtered for galaxies within the redshift range corresponding to $38.6-48.6$ Mpc and within a projected distance of $35\arcmin$ ($440$ kpc). With these parameters we identified five galaxies that have K-band magnitudes in the range of $m_{\rm K} = 13.4-15.5$ magnitudes. Therefore, we assume that these galaxies are located within the virial radius of NGC~6753 and compute their stellar mass from their \textit{2MASS} K-band magnitude and the corresponding mass-to-light ratios. The total stellar mass of these satellites is  $M_{\rm \star}  = 1.0\times10^{11}  \ \rm{M_{\odot}}$. Taking these components together, we find that the total baryon mass, excluding the hot gaseous phase, is  $4.4\times10^{11}  \ \rm{M_{\odot}}$. To obtain the total baryon mass within the virial radius of NGC~6753, we add the total hot gas mass. We thus find that the total baryon mass of NGC~6753 is $M_{\rm{b,tot}}=8.4\times10^{11}  \ \rm{M_{\odot}}$. 
Based on the total baryon mass, we compute the baryon mass fraction as $f_{\rm{b}}=M_{\rm{b,tot}}/(M_{\rm{DM}}+M_{\rm{b,tot}})$ and obtain $f_{\rm{b}}= 0.06 $. Given that the cosmic value is $f_{\rm{b,WMAP}}=0.156\pm0.002$ based on the measurements from \citet{planck16}, we conclude that NGC~6753 is missing about half of its baryons (Figure \ref{fig:fgas}). This conclusion is in line with other studies that established that even galaxy groups, which may have similar or slightly higher virial mass, have a fraction of their baryons missing \citep{giodini09}.

In addition to the baryon mass fraction within the virial radius, we also derive the baryon mass fraction within 50 kpc of NGC~6753. We note that the baryon mass fraction within this region is not affected by systematic uncertainties due to the unknown gas density profile at larger radii. While this measurement is less representative, due to its robust nature it provides a useful basis to compare with the results of present and future galaxy formation simulations. To derive the baryon mass within this radius, we include all of the above listed baryonic components except for the stellar mass of other galaxies. Within 50 kpc the total hot gas mass is $1.0\times10^{10}  \ \rm{M_{\odot}}$, hence the total baryon mass is $3.3\times10^{11}  \ \rm{M_{\odot}}$. Note that the total mass within this region is dominated by stars. The total dark matter halo mass within this region is inferred from Section \ref{sec:hydro} and is $1.7\times10^{12}  \ \rm{M_{\odot}}$. Hence, the total baryon mass fraction is $f_{\rm b} \approx 0.16$, which is in line with the Planck results (Figure \ref{fig:fgas}).

\subsection{Metallicity of the gas}
\label{sec:metals}

The spectral analysis of the hot gas noted that it has strictly sub-Solar ($Z\sim0.1$ Solar) abundance at every radii. These values are similar albeit slightly lower than the values obtained for NGC~1961, which showed a metallicities consistent with $Z\sim0.2$ Solar at every radius \citep{anderson16}.  The low metallicities for NGC~1961 and NGC~6753 are in stark contrast with the values obtained for luminous elliptical galaxies. Indeed, in massive ellipticals the metallicity of the hot gas is around Solar \citep{kim04,humphrey06}, while lower-mass ellipticals ($\lesssim10^{11} \ \rm{M_{\odot}}$) exhibit sub-Solar metallicities \citep{bogdan12} similar to that observed for NGC~6753 or NGC~1961. 

As a caveat we mention that the low metallicities may partly be caused by the so-called iron-bias \citep[e.g.][]{buote00}. In Section \ref{sec:metallicity} we demonstrated that the hot gas in the corona of NGC~6753 exhibits a temperature gradient and an azimuthal structure. Based on these, it is feasible that applying a single temperature model to measure the metal abundances results in an underestimated metallicity. To alleviate this issue, it would be ideal to utilize a more complex model, including two thermal components. However, in the outskirts of  NGC~6753 the signal-to-noise ratios are not sufficiently high enough to utilize a two temperature model and obtain constraining parameters. Thus, based on the present data set it cannot be excluded that the true metallicities are significantly higher than those of the best-fit values obtained from fitting single temperature models. 

Based on the hot gas mass within 50 kpc ($1.0\times10^{10}  \ \rm{M_{\odot}}$) and the virial radius ($4.0\times10^{11}  \ \rm{M_{\odot}}$) and by using the metallicity of 0.1 Solar, we estimate the total iron mass in NGC~6753. Given the iron abundance of $1.23\times10^{-3}$ by mass \citep{grevesse98}, we estimate that the total iron mass within 50 kpc and within the virial radius is $1.23\times10^{6}  \ \rm{M_{\odot}}$ and $4.92\times10^{7}  \ \rm{M_{\odot}}$, respectively. Type Ia Supernovae (SN Ia) play a major role in enriching the interstellar medium with iron-peak elements. Specifically, each SN Ia event provides about $0.7 \ \rm{M_{\odot}}$ of iron \citep{nomoto84,iwamoto99}. For S0a/b galaxies the typical SN Ia rate is $0.046^{+0.019}_{-0.017}$ SNuM, where 1 SNuM is equivalent with 1 SN Ia per $10^{10} \ \rm{M_{\odot}}$ per century \citep{mannucci05}.  Given the stellar mass of NGC~6753, this corresponds to an SN Ia rate of $0.021 \ \rm{year^{-1}}$ and an iron yield of $0.015 \ \rm{M_{\odot} \ year^{-1}}$.  Therefore, the iron mass within 50 kpc and the virial radius can be produced in about 82 Myrs and 3.3 Gyrs, which timescales are much shorter than the age of the galaxy. This implies that a significant fraction of the iron produced by SN Ia may be missing, especially from the innermost regions of NGC~6753. It is feasible that at least part of  the iron does not mix with the interstellar medium \citep{brighenti05}, but cools to low temperatures, and hence remains hidden from X-ray observatories. Alternatively, part of the iron may leave the galaxy in a wind to the outskirts of the corona. This would imply that the iron abundance is not uniform out to the virial radius, but a metallicity gradient may be present. Although  it would be crucial to probe the outskirts of the hot corona around NGC~6753 to resolve these outstanding questions, due to the faint nature of the coronae this remains infeasible for the current generation of X-ray telescopes (Section \ref{sec:metals}). 

In a simple picture, low metallicity can be achieved by either metal-rich outflows or metal-poor inflows. Given that the metallicity of NGC~6753 and NGC~1961 is consistent with $\sim0.1-0.2$ Solar out to $\sim50$~kpc, it is unlikely that metal-rich outflows dominate the hot corona of these galaxies. Heavy elements lifted by the outflows would mix with the hot gas and result in a positive metallicity gradient outwards, which is inconsistent with our \textit{XMM-Newton} data. Therefore, in (massive) spiral galaxies metal-poor inflows of primordial gas likely play a role. As opposed to this, hot gas arising from the stellar yields of evolved stars and heated by the energy input of supernovae may play a more notable role in elliptical galaxies \citep{knapp92,bogdan08,conroy14}. The reality, however, is possibly more complex than this simple picture, since multiple processes may play a notable role, such as simultaneous inflow and outflow, mixing and stirring of the gas in the corona with the gas expelled from winds, and AGN activity. To further probe this picture, it would be essential to confront our observational results with modern galaxy formation simulations, such as IllustrisTNG \citep{vogelsberger17}. However, such a comparison is beyond the scope of the present work, and will be the subject of a future publication.

\subsection{Future prospects}
\label{sec:metals}

The present study of NGC~6753 represents the second detailed investigation of the hot X-ray corona around a massive spiral galaxy. While the main conclusions of the properties of the hot corona are similar for NGC~6753 and NGC~1961, it must be realized that both the present work and the study by \citet{anderson16} explored the hot gas residing in the inner $\approx0.11r_{\rm 200}$ of two extremely massive spiral galaxies. Therefore, to generalize the conclusions achieved in these works, it  would be necessary to explore the hot corona out to larger radii and to probe a larger sample of less massive galaxies. However,  as we discuss below, due to the limitations of the currently operating X-ray missions, the exploration space is limited. 

Given that the dominant fraction of the hot gas resides in the outskirts (i.e. beyond $0.11r_{\rm 200}$) of the galaxies, it would be essential to observationally probe the characteristics of the hot gas at these larger radii. However, due to the low density and/or the low temperature of the gas, the emission measure of the gas drops rapidly with an increasing radius. As a result, the systematic uncertainties associated with the relatively high background level dominate at large radii, and make any studies virtually unfeasible.

Since NGC~6753 and NGC~1961 represent the rare class of massive star-forming galaxies, it would be ideal to explore the hot gaseous coronae of lower-mass spiral galaxies. However, galaxy formation simulations predict that the hot corona around these galaxies is significantly fainter and has lower gas temperature. This is consistent with the non-detections of X-ray coronae around low-mass spiral galaxies \citep{bogdan15}. Due to the relatively low sensitivity of \textit{XMM-Newton} and \textit{Chandra} at energies below $0.5$ keV and the associated high background, it is extremely challenging to detect the extended X-ray  corona around Milky Way type galaxies with these instruments. 

To explore the hot gaseous coronae out to larger radii and around lower-mass spiral galaxies, the next generation of telescopes is required. Indeed, \textit{Athena} or the proposed \textit{Lynx} mission will provide the necessary collecting area to study the extended hot corona around a notable sample of spiral galaxies out to a significant fraction of the virial radius. The observations taken with the next generation of X-ray telescopes will allow us to further probe the temperature and metallicity structure of the gas, thereby understanding the evolutionary processes that influence the galaxies from their formation until the present day.

\section{Conclusions}
In this work, we study the hot gaseous X-ray corona around NGC~6753, a massive spiral galaxy. To this end, we utilize $290$~ks deep \textit{XMM-Newton} observations. Thanks to these deep observations we can explore the gaseous corona of the galaxy in unprecedented detail. Our results can be summarized as follows.

\begin{enumerate}
\item We confirm the presence of a luminous X-ray corona around NGC~6753, which can be robustly traced out to $50$ kpc radius from the nucleus of the galaxy. Beyond this radius the systematic uncertainties dominate the overall X-ray signal. 
\item Based on a denoised X-ray image we find that the gaseous X-ray emission is isotropic and does not show signatures of disturbances or outflows. 
\item We derive a detailed gas temperature profile that reveals a decreasing trend from $0.75$ keV in the central regions to $0.4$ keV beyond $\sim25$~kpc. We further probe the temperature structure of the gas by constructing a temperature map of the inner 30 kpc of the galaxy, which demonstrates the complex temperature distribution.
\item We explore the metallicity distribution of the hot gas and find that it shows strictly sub-Solar ($Z\sim0.1$ Solar) metallicities at every radius. We speculate that the low metallicities may be a consequence of metal poor cold flows and the quiescent star-formation history of NGC~6753. 
\item By extrapolating the gas mass profile, we estimate the total baryon mass of NGC~6753, and compute the baryon mass fraction of the galaxy. We conclude that it is $f_{\rm{b}}= 0.06 $, which falls short of the cosmic value, implying that the galaxy is missing about half of its baryons. \\

\end{enumerate}

\bigskip

\begin{small}
\noindent
\textit{Acknowledgements.}
We thank the referee for constructive comments. This work uses observations obtained with \textit{XMM-Newton}, an ESA science mission with instruments and contributions directly funded by ESA Member States and NASA. This publication makes use of data products from the Two Micron All Sky Survey, which is a joint project of the University of Massachusetts and the Infrared Processing and Analysis Center/California Institute of Technology, funded by the National Aeronautics and Space Administration and the National Science Foundation.  In this work the NASA/IPAC Extragalactic Database (NED) have been used. We acknowledge the usage of the HyperLeda database (http://leda.univ-lyon1.fr). \'A.B., W.R.F, and R.P.K.  acknowledge support from the Smithsonian Institution. 
\end{small}

\end{document}